# Nonequilibrium dynamics of spontaneous symmetry breaking into a hidden state of charge-density wave

Faran Zhou[1], Joseph Williams[1], Christos D. Malliakas[2,3], Mercouri G. Kanatzidis[2,3], Alexander F. Kemper[4], Chong-Yu Ruan[1*]

[1] Department of Physics and Astronomy, Michigan State University, East Lansing, MI 48824, USA.
[2] Department of Chemistry, Northwestern University, Evanston, IL 60208, USA.
[3] Materials Science Division, Argonne National Laboratory, Argonne, IL 60439, USA.
[4] Department of Physics, North Carolina State University, Raleigh, NC 27695, USA
* e-mail: ruan@pa.msu.edu

Nonequilibrium phase transition plays a pivotal role in a broad physical context from condensed matter to cosmology. Tracking the formation of non-equilibrium phases in condensed matter is challenging and requires a resolution of the long-range cooperativity on ultra-short timescales. Here, we study the spontaneous symmetry breaking transformation of a charge-density wave system from a stripe phase into a checkerboard state. Such a state is thermodynamically forbidden, but is introduced through a system quench using ultrashort, intense laser pulses. The dynamics is mediated by the soft modes that unfold spontaneously and order the field on a timescale ≈1 ps. Using the coherent electron diffraction with ≈100 fs resolution, we capture the entire course and demonstrate nonergodic behavior proximal to symmetry breaking that is crucial for stabilizing the hidden states. Remarkably, the thermalization due to carriers cooling arrests the remnants of the transient orders into the topological defects in the eventual state with distinct new properties that last for more than 1 ns. The fundamental dynamics observed here opens an intriguing perspective of controlling phase transitions in quantum materials far from equilibrium.

The remarkable feature associated with spontaneous symmetry breaking (SSB) is emergent scale-invariant dynamics in approaching a thermal critical point[1]. There have been strong incentives to understand how this self-organization may proceed out of thermal equilibrium[1,2]. Studying the nonequilibrium phase transition introduced via a swift change of the system parameters, also called a quench, is one of the most active areas in nonequilibrium physics, impacting diverse fields from condensed matter[3,4], quantum gases[2,5,6], to cosmology[4,7]. It is widely believed that, after an interaction quench, isolated systems generically approach a thermal state[2]; however, a transient nonthermal stationary state may emerge with properties unlike their equilibrium counterparts[8]. Such investigations have been carried out using ultracold atoms[2,5]. Meanwhile, recently ultrafast pump-probe studies made surprising discoveries of light-induced superconductivity[9] and insulator-metal transitions in hidden charge density wave (CDW) states[10,11], hinting undisclosed routes towards new symmetry-broken states[12]. Extending the delicately controlled quantum gases experiments to the condensed matter sector is a crucial yet challenging task.

In this letter, we demonstrate for the first time that non-equilibrium crossing an interaction-mediated critical point introduces an entirely unexpected CDW order. The system is CeTe$_3$[13] and the interaction quench is engineered via the femtosecond (fs) infrared pulse excitation which alters the system preference of SSB from the stripe order to a bi-directional state not permitted in a thermal state[14,15]. This phase transition occurs exclusively on nonequilibrium timescales (100 fs-1 ps), and it displays key dynamical characteristics of a quenched SSB system: the spontaneous emergence of soft-mode instabilities, followed by coarsening to adopt the new broken-symmetry. Critical for studying this far-from-equilibrium self-organization is a major advance in fs scattering through generating intense and coherent electron pulses [16,17], enabled in a prototype ultrafast electron microscope column (Fig. 1a). This new setup employs adaptive optical control[17] to deliver very high beam brightness, translated into $\geq 10^3$ increase in flux and $\geq 10$ improvement in beam coherence length[11,18], allowing us to capture the unfolding of the critical events with atomic details (see Methods Sec. 1).

To aid the understanding of this unexpected process, we describe the non-equilibrium phase transition using a conceptual Landau-Ginzburg free energy[1] for the CDW order parameter $\Psi = \Delta e^{-i\phi}$, with $\Delta$ and $\phi$ representing the amplitude and the phase of the field[19]; see Fig. 1b. The quench is mediated by the fs laser pulse carrier excitations, which change the (orbital) interaction control parameter $x$ and swiftly drive unfolding of the energy landscape across the critical point $x_c$. However, the system cannot order spontaneously[3] because the long-wavelength amplitude and phase modes order the order parameter field on much longer timescales[20,21]. Therefore, the ultrafast dynamics of the order parameter reflects the inherent broken ergodicity in a spontaneous symmetry-breaking phase transition[3,21].

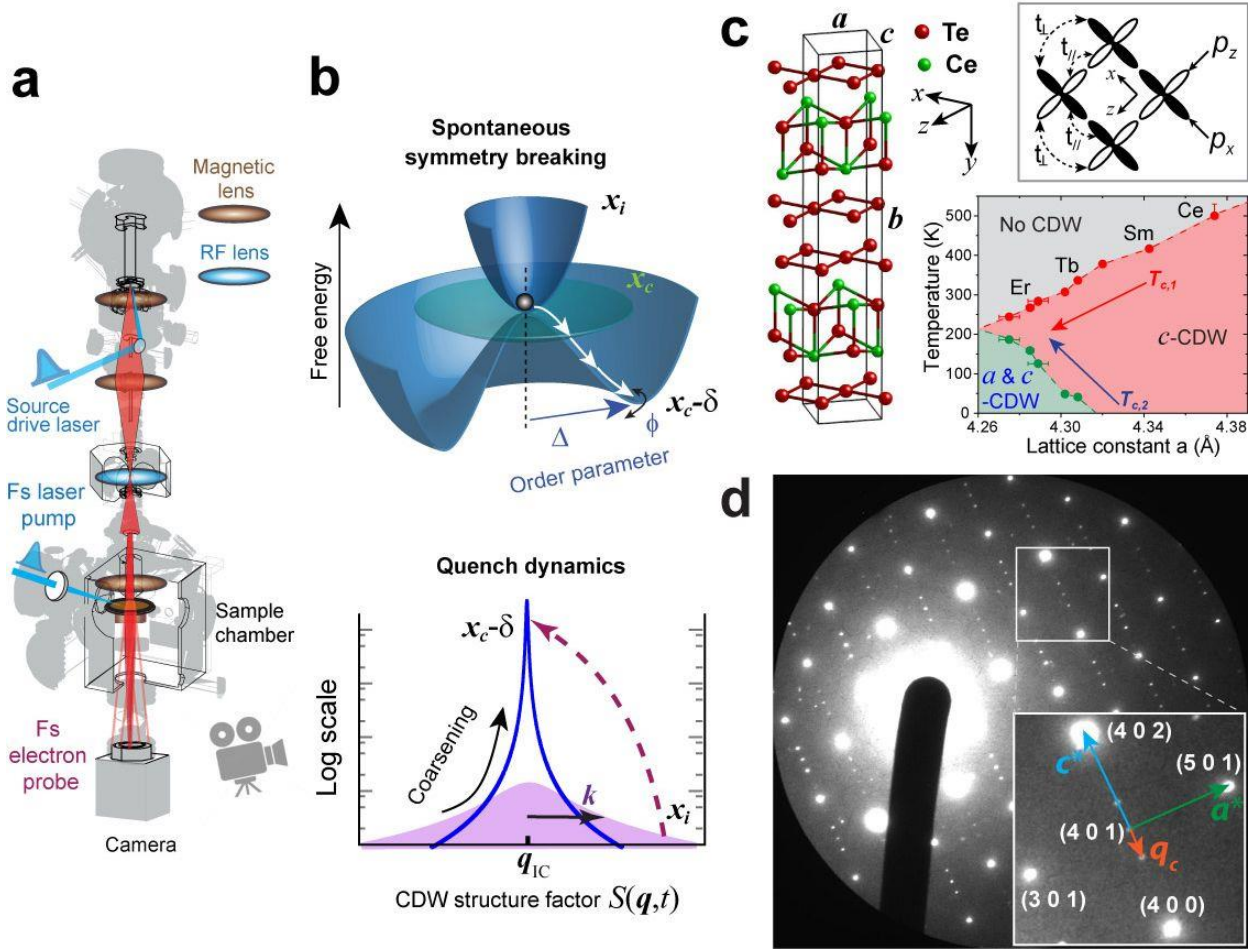

**Fig. 1. Spontaneous symmetry breaking in CeTe$_3$** | **a,** Ultrafast electron microscope column for probing the CDW phase transitions. **b,** SSB process probed using ultrafast electron scattering. The upper panel shows the Landau free energy depiction of an adiabatic SSB process. The lower panel shows the scattering profile changes after a rapid quench. **c,** Structure of CeTe$_3$. Corrugated CeTe layers are sandwiched by square Te nets, in which CDW forms. Inside box depicts the Te 5$p_x$ and 5$p_z$ orbital couplings near the Fermi surface in a unit cell. Inset shows the RTe$_3$ phase diagram with two types of SSB [14,15]. **d,** Diffraction pattern of CeTe$_3$ obtained with the fs coherent electron beam line. Inset shows the CDW satellite peaks located at wave vector $q_c$ away from the main lattice Bragg peak.

Studying how a system adopts long-range correlations after a quench, often called coarsening[1], is a key problem in non-equilibrium physics[3-5,7,20]. While in the late stage of coarsening the system is expected to follow universal laws[1,3], at the shortest time the evolution is dominated by the spinodal instabilities from unfolding of the uphill potential (Fig. 1b)[20]. In a non-equilibrium SSB of an incommensurate (IC) CDW system, the soft modes appear as unstable order parameter fields. In momentum space, they are described as long-wavelength instabilities with $k=q-q_{IC}$ [22], which can be investigated via scattering weight changes around the emergent wave-vector $q_{IC}$[19,21-23] using the coherent fs scattering setup (see Fig. 1b). The correlation function $S(r,t)$ of the developing unstable order parameter fields[1,3,20,21] can be deduced via inverting the structure factor $S(q,t)$ – the Fourier counterpart of $S(r,t)$ – that is studied directly by the scattering intensity $I(\boldsymbol{q}, t)$ at $\boldsymbol{q}_{IC}$. The coarsening, characterized by an increase of the correlation length $\xi(t)$ as the non-equilibrium system unfolds[1,3], is witnessed in sharpening of the scattering peaks[24]. Meanwhile, the amplitude of the order parameter field ($\Delta$) is obtained through integration over momenta: $I_{int}(t) \sim \Delta^2(t)$ (see Methods Sec. 2).

The CeTe$_3$ studied here[13] belongs to the rare-earth tritelluride (RTe$_3$) family[14,25] where the incommensurate CDW develops inside the double Te square lattice sheets, which are isolated by the buckled insulating CeTe layer (Fig. 1c). The RTe$_3$ family are ideal systems for studying SSB because in the square net of Te layers where two types of density waves (stripes and checkerboard) are predicted[26], subjecting to the interplay between the nesting in the electronic structure[25,26] and the strong momentum-dependent electron-phonon coupling (EPC) [27,28]. Within the Te layer, the shape of the 2D metallic Fermi Surface (FS) depends on the relative coupling strengths between neighboring $5p_x$ and $5p_z$ orbitals ($t_\perp$ and $t_\parallel$ in Fig. 1c) [25,26]. The predominant CDW state is $c$-CDW, chosen by a subtle bi-layer coupling that breaks the C4 symmetry[26]. In the heavier member CeTe$_3$, a large $c$-CDW gap removes a significant amount of the potential $a$-CDW spectral weight [25,26], which completely excludes subsequent formation of $a$-CDW in its equilibrium phase diagram (Fig. 1c)[15].

Our CeTe$_3$ sample is a single crystal exfoliated to a thickness of ≈25 nm to match the pump laser penetration depth (see Methods Sec. 1). The distinctly uniaxial CDW formation prior to laser excitation is identified in our fs diffraction setup. The diffraction pattern (Fig. 1d) shows sharp $c$-CDW satellite peaks at $\boldsymbol{q}_c \approx 2/7\boldsymbol{c}*$ [13,14] on both sides of the lattice Bragg peaks ($\boldsymbol{G}$). To contrast the laser-induced transformations, we first study the thermal phase transition with a TEM; see Supplementary Data Fig. 1 for the results. We observe CDW melting to occur at $T_c \approx 540$ K. Prior to transitioning into the normal state at $T_c$, the wave-vector $\boldsymbol{q}_c$ does exhibit a steady shift starting from 400 K[14]. Yet, the CDW system maintains a large static correlation length ≈40 nm up to 480 K.

The nonthermal phase transition is driven by 800 nm, 50 fs laser pulses. To target nonthermal melting, we employ a pump fluence ($F$) of 1.86 mJ/cm$^2$ capable of instantaneous carrier heating to a temperature ≈ 5,000K (see Methods Sec. 4). At this excitation level, the quasiparticle gap of the CDW system collapses within ≈250 fs as demonstrated by the ultrafast optical and angle-resolved photoemission spectroscopy (ARPES) experiments[29-32]. Uniquely here, the corresponding dynamics represented in the lattice counterpart[14,23,25] can be followed by the fs coherent electron scattering. Given that the excitation of the carriers occurs before the two systems (carrier and lattice) have time to equilibrate, it can influence the system parameters such as FS nesting and EPC that promote the initial CDW ordering[27,28,30] (see Methods Sec. 5) and steers the system to different types of broken symmetries[26]. The key questions are: at what stage does the symmetry breaking (or recovery) occur, and to answer the key question here, can the system reorder without reaching thermal equilibrium?

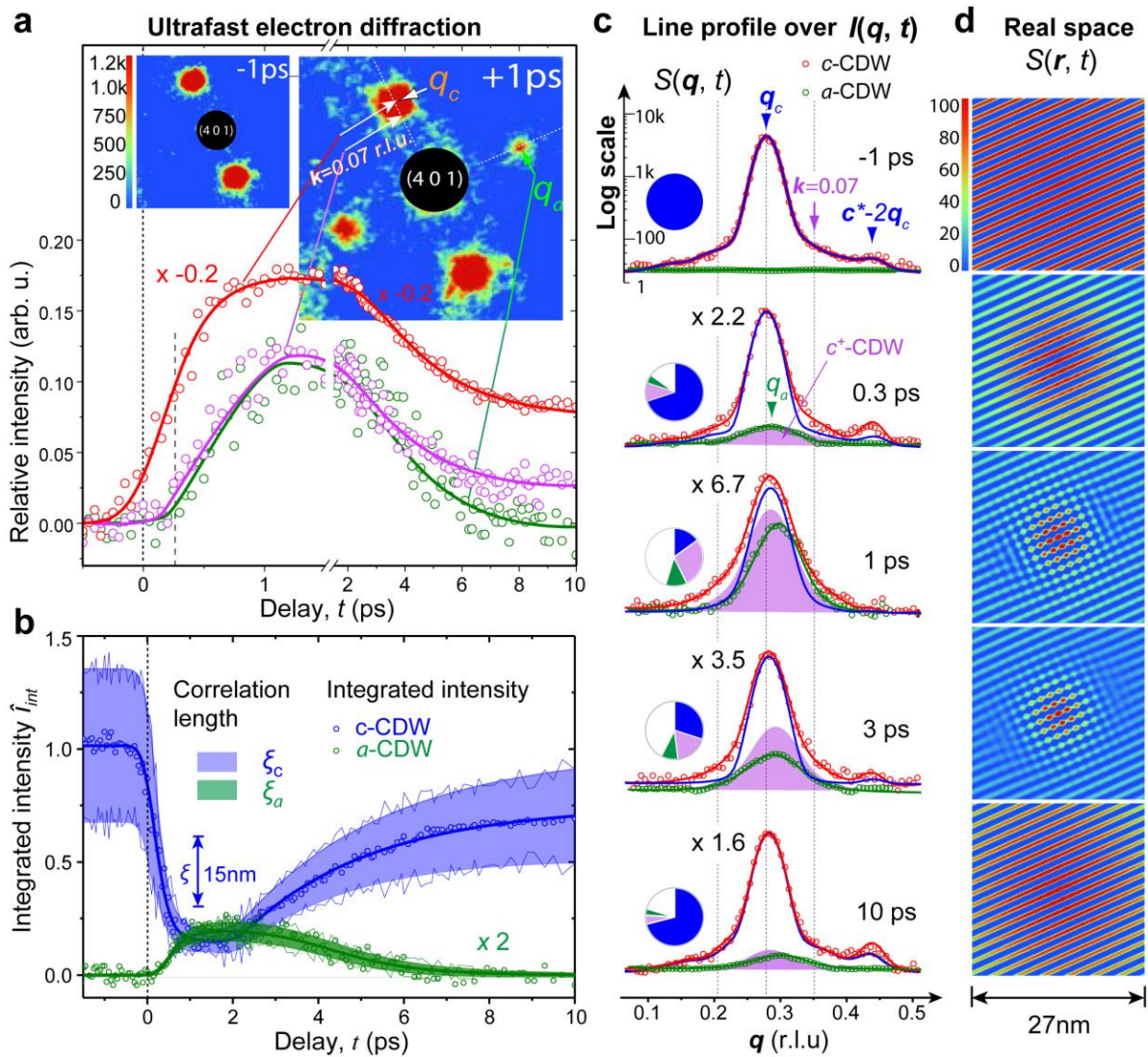

**Fig. 2. Nonequilibrium CDW phase transitions** | **a,** Scattering weight transfer in the ultrafast electron diffraction near a nonthermal critical point. The scattering dynamics representing each CDW channel are color-coded: static $c$-CDW (blue), emerging $c^+$-CDW (purple), and $a$-CDW (green). Top inset images show the diffraction patterns before and after laser quench with clear emergence of $a$-CDW satellite at $q_a$ while the preeminent $c$-CDW peak loses strength. The panel shows the relative scattering weight evolution between the decreasing scattering from the static order and the increasing diffusive scattering from the $c^+$ and $a$-CDWs. **b,** The panel shows the comparison between the evolution of the static $c$-CDW and the emerging $a$-CDW based on integrated intensity. The correlation length ($\xi$) of each state is also presented in the shaded envelop. **c,** Temporal evolution of the CDW profiles taken along $a$- and $c$-CDW satellites. The data symbols are fitted with the theoretical diffraction profiles, calculated (through Fourier Transform) from the CDW correlation functions, $S(r,t)$, see Methods Sec. 2. The pie charts (same color codes as in **b**) indicate the relative portion of the CDW components during evolution. **d,** The corresponding $S(r,t)$ constructed based on the refinements using the parameters deduced via the fs diffraction.

We present our key results of the fs diffraction experiments in Fig. 2. Upon applying the nominal fluence laser quench, the subtle melting and re-ordering dynamics are captured by the momentum-dependent features; see Fig. 2a. Here, we focus on the structure factor [$S(q,t)$] near the (401) Bragg peak, where the initial $c$-CDW order parameter undergoing rapid (partial) melting is witnessed from suppressing the intensity to ≈15% in just 300 fs. This is accompanied by strong CDW instabilities reducing the correlation length from >30 nm to ≈ 5 nm shown in Fig. 2b at the same time. These dynamics are consistent with extinguishing the order parameter towards the normal state on fs timescales[19,23]. The inability of the long-wavelength modes to act immediately leaves the initial responses from the static $c$-CDW order parameter dominated by the excitations at short wavelengths[20]. During this period, local defects in the order parameter field will continuously form and get annihilated [5,20,22], reducing the correlation length nearly spontaneously. As the system starts to cool, the defects will coarsen into domains of the different broken-symmetry states until they are unable to do so further[3,4,6,20]. This is shown in the contrasting momentum($k$)-dependent dynamics in Fig. 2a where the $c$-CDW profile amplitude at $k=0$ (red) is spontaneously damped, but the scattering weight is transferred to the incoherent unstable CDW modes first. The competitive CDW order subsequently emerges from coarsening with a clear delay. This is observed in the growth in scattering amplitude at $k=0.07$ r.l.u. (purple), which is transiently coupled to a surprising growth at the orthogonal direction along [100] (green) where the diffusive peaks evolve into distinctive satellites at $|q_a|=0.30$ r.l.u, at 1 ps; see Fig. 2a.

The multi-component scattering profile analyses along both [001] and [100] let us track in details the evolving order parameter fields and their instabilities. Key results are plotted in Fig. 2c in semi-log scale at different delays. The formation of the new bi-directional orders are clearly evidenced at 300 fs, where not only the $a$-CDW (green open circles) is visible, a new branch of $c$-CDW (colored in purple shade and referred to as $c^+$-CDW hereafter) responsible for the intensity increase at $k=0.07$ r.l.u. is also identified. We note that the correlation length of the static $c$-CDW is reduced to ~25% at this time, indicating a large region of disordered field where the new CDW orders could rise. Evidence for coarsening is demonstrated from the clear sharpening and gaining of the intensity in the $a$ and $c^+$-CDW profiles in the subsequent ~1 ps. To visualize the dynamics the correlation functions $S_l(r,t)$ (where $l$ is the index for $a,c,c^+$-branches) are analyzed from an iterative Fourier transform scheme where the respective structure factors $I_l(q,t) \sim S_l(q,t)$ can be reproduced (line profiles in Fig. 3c) by refining the parameters, including the local amplitudes, wave-vectors, and the correlation lengths of the CDW states in both longitudinal and transverse directions. We remark that in Fig. 3d different correlation functions are overlaid as our technique does not resolve the phase directly. However, our diffraction results clearly indicate a coupled ordering form the two CDWs ($a$ and $c^+$) with a strong possibility of jointly forming a checkerboard state. This formation represents a nonthermal hidden state and decays over several ps once the system thermalizes.

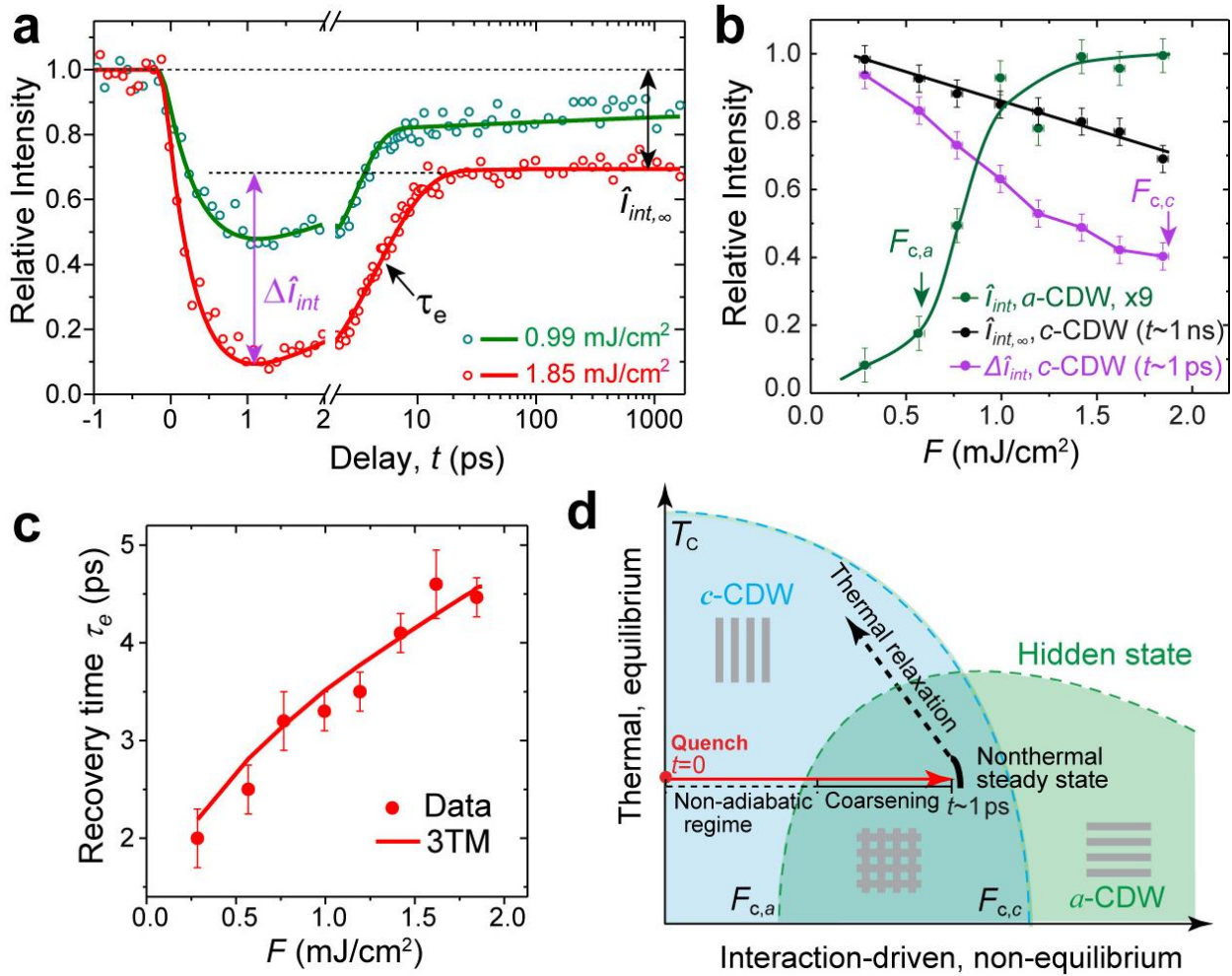

**Fig. 3. Laser parameter space for driving the thermal and nonthermal phase transitions.** | **a,** Common features of $c$-CDW order parameter intensity evolution obtained at two selected laser pump fluences ($F$). The $\Delta \hat{I}_{int}$ in purple represents the nonthermal suppression of the $c$-CDW at ≈1 ps; while $\hat{I}_{int,\infty}$ in black represents the late-state intensity of the thermal state. **b,** Determination of the critical points based on the $F$-dependent CDW scattering intensity extracted from the $a$- and $c$-CDW states. The $F_{c,a}$ and $F_{c,c}$ are determined to be the nonthermal critical thresholds for $a$-CDW formation and $c$-CDW melting. **c,** Recovery time of $c$-CDW state as a function of $F$ and three-temperature model (3TM) predictions. **d,** A phase diagram for the nonadiabatic interaction-driven phase transition revealing the hidden state.

To explore the quench parameter space responsible for creating the hidden bi-directional orders, we examine the thermal and nonthermal characteristics in the evolution as a function of the pump fluence. In Fig. 3a, the common features are shown, depicted in the (normalized) integrated intensity, $\hat{I}_{int}(t)$ at two very different fluences; for the complete data set, see Methods Sec. 4. The dramatic recovery at ≈2 ps marks the decay from the initial non-equilibrium metastable phase to the thermal state that persists into the longer timescale. Within the metastable period (~1 ps), the electronically driven instabilities are evaluated from the over-damped intensity $\Delta\hat{I}_{int}(t\sim 1\text{ ps})$, clearly distinguished from the changes at the late stage $\hat{I}_{int,\infty}(t\sim 1\text{ ns})$ (dashed line) where full system thermalization occurs. These intensity differences are plotted in Fig. 3b, along with the intensity from the $a$-CDW state [$\hat{I}_{int,a}(t\sim 1\text{ ps})$], as a function of $F$. We remark that the recovery time $\tau_e$, which coincides with decay of the $a$-CDW state, is $F$-dependent and is reminiscent of the electron-phonon coupling time identified in the earlier ultrafast ARPES and diffraction experiments[18,29,31-33].

To understand the relaxational dynamics after initial laser excitations[18,29,32], Tao, Han et al. constructed a three-temperature model (3TM)[18,33] to model the evolutions of the local temperatures that they prescribe to the three microscopic systems: the electrons ($T_e$), high-energy phonons that strongly couple to the CDW state ($T_{CDW, phonons}$), and the rest of the lattice modes ($T_{lattice, phonons}$) to capture the observed different time scales[32,33]. The relaxational part of the CDW dynamics starts with the decay of metastable state, which corresponds to the thermalization stage in 3TM, where relaxation of the hot carriers eventually leads to heating the lattice (phonons) bath. The 3TM is reproduced here for our experimental conditions (see Methods Sec. 6). We find the recovery time ($\tau_e$) of the $c$-CDW, as depicted in Fig. 3c, can indeed be captured by 3TM (solid line).

One of most remarkable results to understand hidden state formation lies in its relatively low-energy onset, characterized by $F_{c,a}$ in Fig. 3b, and the fact that the order parameter of the hidden state increases as one increases the pump fluence – in a process contrary to the conventional wisdom of photothermal quenches that typically destroy orders. This directly shows that $F_{c,a}$ is not the conventional thermal critical point, but rather one mediated by non-temperature-associated interaction change. In fact, the critical threshold identified for forming the $a$-CDW ($F_{c,a}$=0.6 mJ/cm$^2$) is more than three and six times smaller than what required to electronically or thermodynamically melt the $c$-CDW state ($F_{c,c}$ =1.9 mJ/cm$^2$ and $F_{c,T}$=6-7 mJ/cm$^2$ respectively); see Fig. 3b. Given that prethermalization is precondition for creating the hidden states here, a phase diagram that describes the landscape for non-equilibrium phase transitions emerges, as depicted in Fig. 3d.

To see how this interaction-related system parameters may be changed by laser excitation, we draw the relevant conclusion from a recent ARPES experiment[30] where the existing curvature of the FS arc – which is the source of asymmetry – is suppressed from a selective light-induced reduction of the transverse coupling constant[26] ($t_\perp$ in Fig. 1c). Indeed, we observe a consistent blue shift in the wave-vector of the hidden state (Fig. 2c) after excitation that agrees entirely with a recovery of the symmetry in inter-orbital coupling both in the trend and the magnitude (see Methods Sec. 5). This establishes the $F_{c,a}$ as an (orbital) interaction-mediated critical point[11,26].

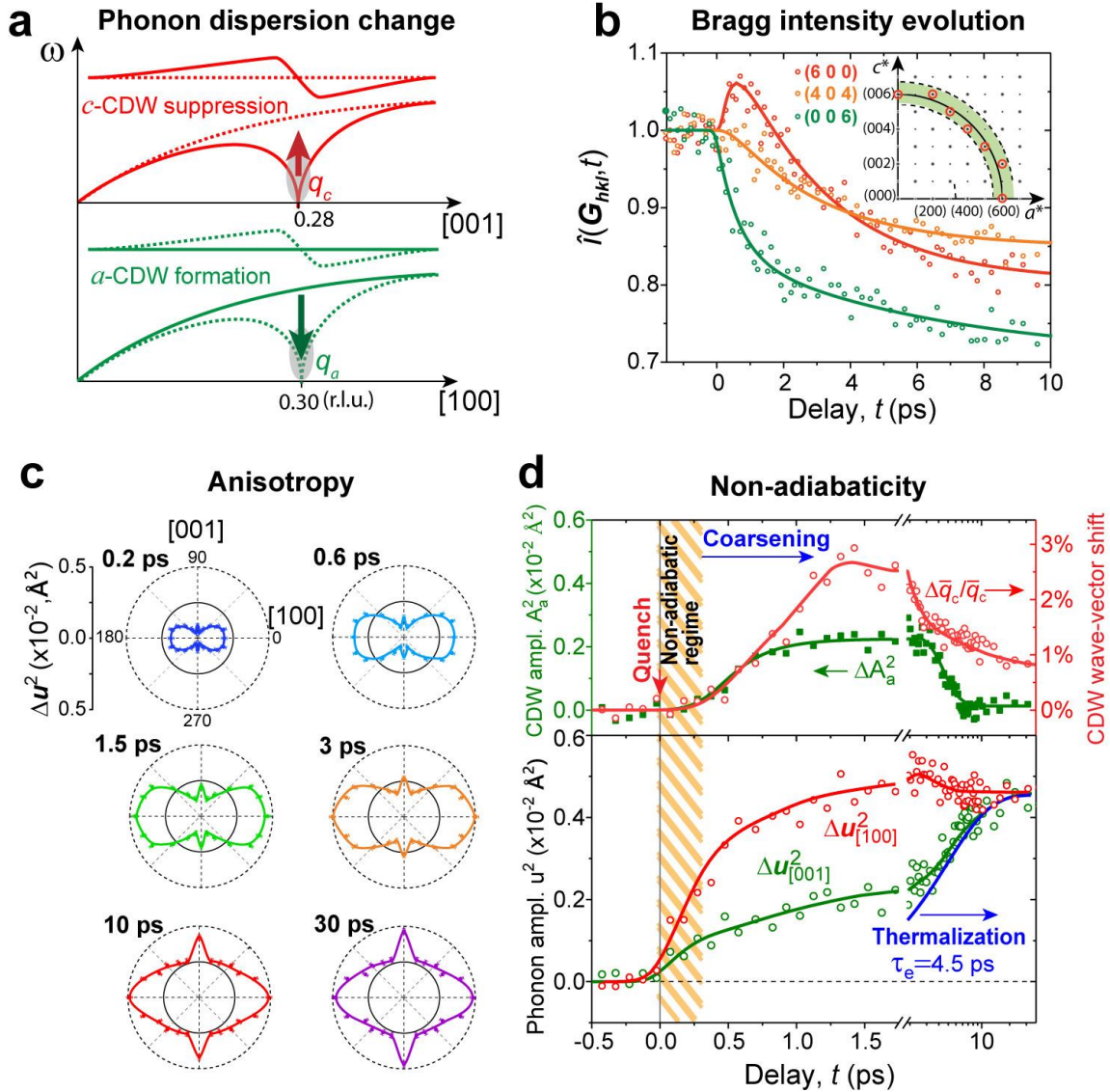

**Fig. 4. Symmetry breaking and recovery energy landscapes and soft modes dynamics.** | **a,** The anticipated phonon dispersion curves over the symmetry recovery (c-CDW) and breaking (a-CDW) processes. **b,** The lattice responses of 3 representative lattice Bragg peaks. Peak intensities obtained along each direction response quite differently. **c,** The mean-squared (ms) phonon amplitude changes obtained using Bragg reflections at $|G_{hkl}|\sim 6$ r.l.u. (see **b**) shown in polar coordinates, indicating strong anisotropy. **d,** (Top) The order parameter field evolution examined via $\Delta A_a^2$ and $\Delta \bar{q}_c/\bar{q}_c$ respectively for a- and c-CDWs. Here, $\bar{q}_c$ is the mean wavevector including $c^-$ and $c^+$ contributions. (Bottom) The vibrational ms phonon amplitude changes projected along [001] and [100] obtained from the Debye-Waller analysis.

The non-equilibrium spontaneous symmetry breaking process may be directly studied in our experiments by following the soft modes. In the laser-induced formation of hidden state, the $c$-CDW is suppressed first and that leads to unstable CDW fields which can adapt into the new broken symmetry of the $a$-CDW state. In the process, the soft modes appear as the unstable CDW fields taking on different polarizations near $q_c$ and $q_a$ as outlined in the changes of the phonon dispersion curves for the two CDWs; see Fig. 4a[19,23]. We study the dynamics of the soft modes based on the lattice mean-squared displacement $u^2_{[hkl]}(t)$ along $[hkl]$, which is resolved by performing the $q$-dependent Debye-Waller analysis using the intensity of the Bragg peaks at $|G_{hkl}| \approx 6$ r.l.u., as depicted in Fig. 4b (see Methods Sec. 7). Both the static (due to CDW formation) and the dynamic (due to phonons including the soft modes) lattice distortions can modify the intensity of the Bragg peaks, but the two must observe the same (broken) symmetry in an SSB phase transition. A key observation is the initial rapid rise of the intensity at (600) [not (006)] in Fig. 4b, which can only occur from a symmetry recovery process taken by the $c$-CDW state at the same period. This points to a different (perpendicular) polarization of the displacement field from the CDW ordering vector, supporting that the soft modes in CeTe$_3$ is a transverse mode[23].

Guided by this understanding, we examine the evolution of soft modes by following post-quench changes in the phonon amplitude $[\Delta u^2_{[hkl]}(t)]$, presented in Fig. 4c in polar coordinates. The soft-mode signatures are clearly noticeable from the strongly peaked amplitude changes along [100] and [001]. In particular, in the first 500 fs the phonon dynamics is dominated by the soft modes from unfolding of the energy landscape. In the bottom panel of Fig. 4d we plot the phonon amplitude changes projected along [100] and [001] to compare with the evolution of the order parameter fields, represented by $\Delta A_a^2$ and $\Delta \bar{q}_c / \bar{q}_c$ respectively for the $a$- and $c^+$-CDW evolutions (top panel). Two features relevant to understand the non-equilibrium SSB stand out here. Firstly, a clear delay is observed in the onset of coarsening to grow the order parameters – a process takes ~ 1ps (upper panel) – relative to the appearance of the soft-mode instabilities (bottom panel). This directly shows that the new CDW order cannot order spontaneously – a hallmark of SSB through a quench[4]. This delay in the subsequent coarsening is indeed recently validated in the simulator experiment using cold quantum gases[5]. Secondly, during thermalization between 2-6 ps the soft-mode amplitudes associated with the $a$-CDW state increase expeditiously; whereas the soft-mode amplitudes from the $c$-CDW state decrease at the same time. This reflects the interconversions between the thermal and photo-induced orders in the two systems.

For the $c$-CDW state, this thermalization dynamics represents cooling of soft modes which drives the system back to the initial stripe order. Intriguingly, the recovered thermal state is unable to fully regain its inherent correlation length (Fig. 2b, >30 nm). This unexpected result indicates that the thermal state may still harbor the remnants of the hidden states which represent the topological defects in the system[4,7,20].

Based on the observed $\xi$ (20 nm), we deduce a defect density of $\approx 2.5\times10^{-5}$ nm$^{-2}$. Despite the sparsity, these defects impose major modifications on the system properties. We can identify an anomalously large shift in the mean CDW wave-vector $\bar{q}_c$ ($\approx 1\%$) that is more than 4 times of the change seen in the thermal state entirely unexpected over the rise (75K) in temperature very far from $T_c$; see Methods Sec. 4 and Supplementary Data Fig. 1. This change, which lasts for more than 1 ns, is accompanied by the continued display of the soft-mode anisotropy characteristic of the hidden state (Fig. 4c) – long after its decay! These astounding results demonstrate yet another way of introducing novel properties into the many-body systems via manipulating the phase transitions in the non-equilibrium realm.

Appling ultrafast, intense laser excitations and probing with coherent electron pulses with high spatiotemporal resolutions, our experiments unveil remarkable, rich behavior in non-equilibrium phase transitions. This includes light-induced state with novel crystallography and properties inaccessible thermodynamically[8,12]. Our experiments show this is achieved by exploiting the inherent nonergodicity in an SSB process proximal to a unique nonthermal critical point established on the interaction quench. In theory, such nonthermal interaction-driven pathway[11] is similar to a quantum phase transition [34]. Critical to establish the photoinduced metastability is to shield the growth of light-induced state from the microscopic thermal fluctuations during coarsening. Normally the impulsively induced new phases are stable only in the prethermal time window, but by further exploiting the nonadiabaticity in the relaxation process new properties can be endowed by freezing the topological defects to alter the characters of the ground state. Together these approaches outline new routes to control properties in quantum materials. A natural question emerges here is whether the hidden states themselves can be controlled or extended into a longer timeframe[10-12]. The may involve controlling the coupling between the macroscopic state the microstate thermal fluctuations[3,20], including by harnessing the driven-dissipative dynamics via variable pumping strategies, such as tuning the wavelength and pulse shape[11], or different sample environments, such as tuning the interfaces and temperature[12].

**Acknowledgements**

The authors acknowledge help from K. Sun with the TEM measurements, and the helpful contributions from E.A. Nowadnick in the simulation. We also thank M. Maghrebi for insightful discussions. This work was funded by the U.S. Department of Energy, Grant DE-FG0206ER46309. The experimental facility was supported by U.S. National Science Foundation, Grant DMR 1625181. A.F. Kemper acknowledge support from U.S. National Science Foundation, Grant DMR-1752713. The work conducted at Northwestern University and Argonne National Laboratory was supported by the U.S. Department of Energy, Office of Science, Basic Energy Sciences, Materials Sciences and Engineering Division.

# Methods

## 1.     Experimental setup

The experiments are conducted using 100 keV electron pulses in an ultrafast electron microscope beamline equipped with the adaptive optical controls[16,17], as schematically depicted in Fig. 1a. A key component is a radio-frequency (RF) cavity, synchronized to the pulsed electron beam, as a lens for manipulating the beam longitudinal phase space, and acting together with the magnetic lens system for transversely focusing the beams. The optical setting here is optimized for a high coherence length ($\geq 40$ nm)[17] to resolve the momentum-dependent dynamics from the CDW satellite reflections. This new design delivers intense ultrashort electron pulses (~100 fs) over a very high single-pulse electron number (~$10^6$ e/pulse) to ensure a sufficient signal-to-noise ratio (S/N). The high-intensity pulses allow the beamline to operate at a moderate pump-probe repetition rate at 1 kHz (or below) to allow a completely thermal recovery of the system[16].

The sample preparation is by exfoliation from the bulk single crystal of $CeTe_3$ synthesized using the chemical vapor transport technique[13]. The 25 nm thin and ~30 μm wide specimen flake is supported in a standard TEM grid placed in a temperature-controlled sample holder initially set at the room temperature (RT=298 K) inside the experiment vacuum chamber. The employment of high brightness electron source[17], allowing us to capture the ultrafast, nonequilibrium processes from weak satellites in the ultrathin $CeTe_3$ single crystal.

## 2.     Non-equilibrium CDW scattering profile evolution and modeling

A data analysis program is developed for this study, aiming to refine the key CDW parameters from the ultrafast diffraction experiments to reconstruct the non-equilibrium dynamics of the CDW phase transition. We start from the real-space ($r$) CDW order parameter field $\Psi_l(r,t) = \Delta_l e^{i(q_l \cdot r + \phi_l)}$, where $l$ denotes $a$, $c$, or $c^+$-CDW and $q_l$ and $\phi_l$ are the corresponding wave-vector and phase. The equal time correlation function for the order parameter field can be written:

$$S_l(r,t) = \langle \Psi_l(r,t)\Psi_l(0,0)\rangle, \qquad (M1)$$

where the bracket denotes averaging. Since the probed system size is much larger than the typical CDW correlation length, Eq. (M1) is simplified from a more exact form of the correlation function: $S(x, x', t) = \langle \Psi_l(x,t)\Psi_l(x',t)\rangle$, where we redefine $x'=x+r$ and $S(x, x', t) = S(x - x', t) = S(r, t)$. The instabilities in the CDW field often lead to reductions of the amplitude ($\Delta_l$) and correlation length ($\xi_l$) after ensemble averaging. We remark a key difference between the non-equilibrium critical dynamics and an equilibrium one. At equilibrium, even though $\langle \Psi_l \rangle_{eqn.} = 0$, there will be spontaneous, usually very small fluctuations on a local scale, near the critical point. However, in the non-equilibrium dynamics after by a quench the fluctuations are typically significantly larger. The effective order parameter field is written here to include the correlation length effect, modelled as a domain centered around a reference position $r_0$ :

$$\Psi_l(r,t) = \Delta_l e^{i(q_l(t)\cdot r + \phi_l)} e^{-\frac{|r-r_0|}{2\xi_l(t)}}. \quad (M2)$$

The structure factor $S(q,t)$, effectively the experimental CDW intensity profile $I(q,t)$, is just the Fourier counter part of the correlation function, i.e.:

$$S(q,t) = \int dr\, e^{iq\cdot r} S(r,t). \quad (M3)$$

Since different CDW states occur at different momenta, $\Delta_l$, $q_l$, and $\xi_l$ are determined independently from the same electron diffraction pattern. It is easy to see the correlation length $\xi_l = 1/w_l$, where $w_l$ is the half-width-at-half-maximum of the diffraction profile, and $S(q,t)$ is a Lorentzian. This is obtained by assuming an exponential decay in correlation, namely $S(r,t) = |\Delta|^2 e^{-|r|/\xi}$, To determine $\Delta$, we integrate $S(q,t)$ over momenta near $q_c$, which gives the intensity integral $I_{int}(q,t) = \int dq\, I(q,t) = \int dq \int dx\, e^{iq\cdot x} S(x) = S(0) \sim |\Delta|^2$.

To construct the maps of periodic lattice distortion (PLD), we rely on the Fourier relationship between the individual structure factor of the CDW states and the correlation function $S(r,t) \sim |\Psi(r,t)|^2$. We include $G(q)$, the Gaussian convolution function representing the instrument limit of the detector, and arrive at

$$I_{theo}(q,t) \approx FFT\{S(r,t)\} \otimes G(q), \quad (M4)$$

where FFT represents the Fast Fourier Transform employed in the modeling. $S(r,t)$ is constructed using $\Delta_l$, $q_l$, and $\xi_l$ determined from fitting the individual CDW satellite intensity $I_l(q,t)$ at $q_l$, which is a Voigt function [a convolution of Lorentzian with $G(q)$], in a procedure where $I_{theo}(q,t)$ is compared to the experimental profile. Typically, $G(q)$ is determined in-situ through fitting the neighboring lattice Bragg peak with its diffraction width limited by $G(q)$ due to the much larger correlation length of the lattice. The width of $G(q)$ ($\sigma_G$) here is $\approx 0.018\text{Å}^{-1}$ and is held constant over the refinements. The initial conditions of $\Delta_l$, $q_l$, and $\xi_l$ are obtained using the Voigt function fitting, see Sec. 3. We remark that in the results of $S(r,t)$ presented in Fig. 3d, the three correlation functions for $a$, $c$, and $c^+$-CDWs are simply overlaid as our technique is not sensitive to $\phi_l$.

To translate from the CDW field amplitude to the local atomic displacement, we write the displacement vector:

$$\delta r_i(t) = A_l Re\{\Psi_l(r_i,t)\} e_j, \quad (M5)$$

with $e_j$ representing the polarization direction, set to be perpendicular to $q_l$ (transverse mode). The transient (normalized) intensity integral $\hat{I}_{int,l}(t)$ is obtained using the $c$-CDW intensity integral at the negative time as the normalization factor, where the lattice distortion amplitude $A_c$ =0.15 Å[18]. The PLD map is constructed as:

$$A_l(\boldsymbol{r},t) = \sum_{i,l} Z_i^2 e^{-\frac{(r-r'_i(t))^2}{u_i^2}} e^{-\frac{|r-r_0|}{\xi_l(t)}}, \quad (M6)$$

where $i$ denote each atomic site and $Z_i$, $\boldsymbol{r}_i$ and $u_i$ are the atomic number, equilibrium position, and mean amplitude of vibration. The displaced position $\boldsymbol{r}'_i = \boldsymbol{r}_i + \boldsymbol{\delta r}_i$. The Extended Data Fig. 1 shows the comparison between the theoretical diffraction pattern deduced from the refinement procedure and the experiments.

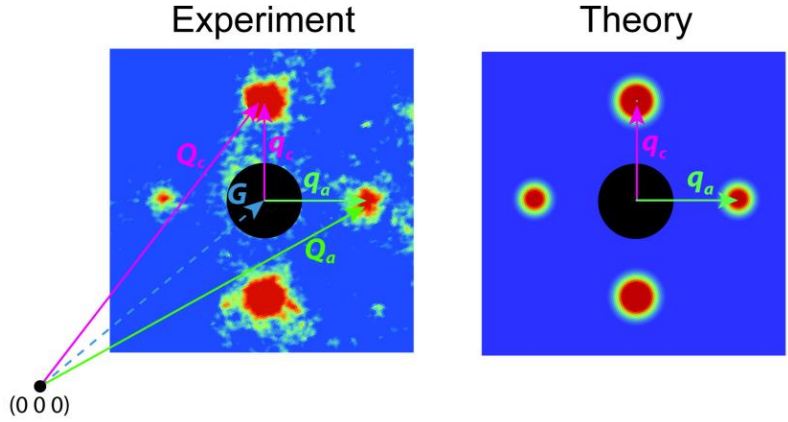

**Extended Data Fig. 1. Theoretical modeling of CDW diffraction pattern** | The scattering pattern is reconstructed based on the Fourier Transform of the real-space correlation function.

For data refinement, we obtain the theoretical diffraction pattern through a Fast Fourier Transform (FFT) followed by a convolution with the instrumental response function $G(\boldsymbol{q})$:

$$I_{theo}(\boldsymbol{q},t) \approx FFT\{I(\boldsymbol{r},t)\} \otimes G(\boldsymbol{q}) \quad (M7)$$

for comparing with the experimental results.

## 3. Transient CDW scattering profile fitting

The first step of the data refinement is to obtain the initial conditions for the input parameters $\Delta_l$, $\boldsymbol{q}_l$, and $\xi_l$, which we obtain by employing Voigt function fitting of the respective CDW satellite intensity profiles (Extended Data Fig.2a). The negative time correlation length is determined to be in excess of 30 nm shown in the Voigt function fitting where the Lorentzian width is extracted by considering a fixed instrumental $G(q)$, shown in Extended Data Fig. 2b. The overall intensity profile at 1 ps (red open circles) cannot be reconciled with the fitting using only one Voigt function. The contribution from the $c^+$-CDW (pink) is isolated by preforming two-Voigt profile fitting shown in Extended Data Fig. 2c. In the fitting procedure, the background is removed via minimizing the non-negativeness with a linear polynomial.

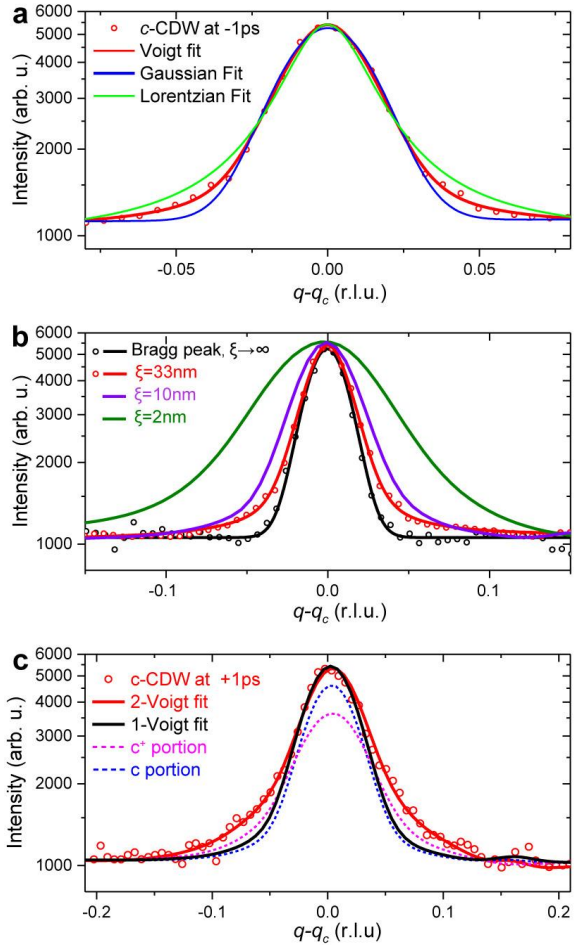

**Extended Data Fig. 2. Peak profile fitting for extracting correlation length $\xi$ | a,** The elastic profile (circle) taken from a $c$-CDW satellite peak at -1ps is best fitted with a Voigt function (red), in comparison with the Gaussian (blue), and Lorentzian (green) functions. **b,** Retrieval of the Lorentz parameters (amplitude and correlation length) through fitting with the Voigt function. The instrumental response is determined from a near-by lattice Bragg peak fitted with a Gaussian (in black). For better comparison, the central amplitudes of the profiles are scaled to the same value. **c,** Intensity profile of $c$-CDW at +1 ps, where a single-Voigt function (black curve) cannot fit the peak well. The two-Voigt function fitting is required to reach a good agreement. Here, the blue dashed line represents the suppressed initial $c$-CDW order and the pink dashed line represents the $c^+$ portion. Red solid line is sum of the two dashed portions.

The second stage of refinement is conducted using the reconstruction protocol in an iterative procedure described in Sec. 2 to reproduce the line profiles of the respective CDW satellite peaks. The results are plotted in Fig. 2c, where the data are in symbols and the theoretical curves are in solid curves. The fits are better constrained by fitting the vertical and transverse profiles at the same time. This better accounts for the diffusive scattering as the background level is similar in both directions. In the data presented in Fig. 2c near the phase transition, the error bars for the peak position is 0.002 r.l.u., and the intensity about 5%.

## 4. Energy thresholds for CDW phase transitions

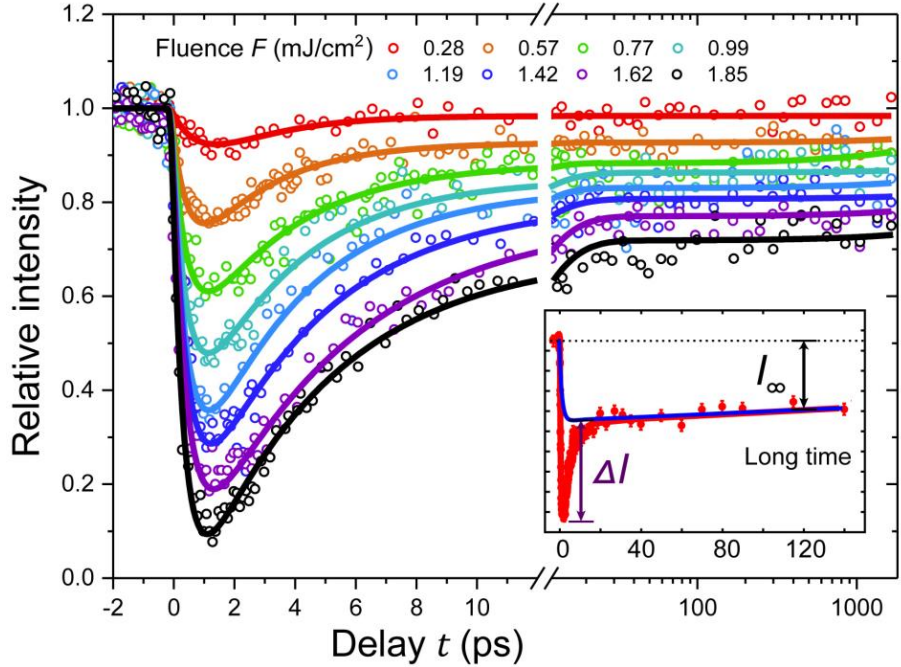

**Extended Data Fig. 3. The fluence-dependent *c*-CDW scattering amplitude evolution** | The inset shows the two different amplitudes representing the nonthermal ($\Delta I$) and thermal ($I_\infty$) responses to the laser excitation.

From the optical studies of CeTe$_3$, the absorbed energy density is calculated to be: $E = F(1-R-T)/[d \cdot (u.c.v.)] = 0.6 \pm 0.2$ eV /(u.c.v.). Here, the unit cell volume[13] (u.c.v.) is 0.503 nm$^{-3}$, and the sample thickness $d=25\pm10$ nm. The reflectivity $R=0.6$, transmissivity $T=0.1$ at 800 nm are determined using the transfer-matrix method[18,35]. Accordingly, we estimate the CDW condensate energy $E_c=1.22$ eV/nm$^3$ using the mean-field formula[36] $E_{el} = n(\epsilon_F)\Delta_e^2[1/2 + ln(2\varepsilon_F/\Delta_e)]^2$. Here, $\Delta_e \approx 0.35$ eV is the CDW gap [37] and $n(\varepsilon_F)=1.48$ state/eV/(u.c.v.) from the ungapped density of states near the Fermi energy, $\varepsilon_F = 3.25$ eV.[25] Meanwhile, to reach the critical temperature $T_C = 450$ K from RT, the threshold for thermal melting $F_{c,T} \approx 7$ mJ/cm$^2$ is obtained based on the specific heat of $\approx 120$ J/(mol·K), assuming the absorbed energy is transferred to the kinetic energy in the system for heating.[38] At the nominal fluence of 1.86 mJ/cm$^2$, the effective temperature $T_{eff}$ of the specimen is $\approx 373$ K (a $\approx 75$ K rise from RT). These results can be compared with the ultrafast measurements. First, the theoretical threshold for thermal melting is consistent with the estimate of 6-7 mJ/cm$^2$ deduced based on the $I_\infty$ in Extended Data Fig. 3; also see Fig. 3b. Second, from the overdamped intensity $\Delta I$ we deduce a threshold of $F_{c,c} \approx 1.9$ mJ/cm$^2$, which gives $E \approx 1.2$ eV/nm$^3$ – a threshold indeed close to the electronic melting threshold $E_c$.

## 5. Symmetry recovery due to laser excitation of $RTe_3$: Theoretical calculations

The possibility of symmetry recovery from pulsed laser excitation in $RTe_3$ is described by Rettig et al.[30] in their photoemission experiment, where the FS curvature is reduced in just 250 fs at an absorbed fluence of 0.27 mJ/cm$^2$ (equivalent to $F$=0.9 mJ/cm$^2$). They mapped this onto a reduction in one of the inter-orbital coupling $t_\perp$ (see Fig. 1c), which in turn is reflected in a reduction of the nesting vector $q_\chi$. A reduction in $t_\perp$ as large as 25% is necessary to account this change. Here, we will show that this reduction leads to the observed change in the CDW wave-vector $q_c$. We evaluate the pairing vector by calculating the static charge fluctuation susceptibility $\chi(\mathbf{q}, \omega = 0)$ based on a model band structure that incorporates the reduction in $t_\perp$. Although $\chi(\mathbf{q}, \omega = 0)$ has peaks at several potential nesting vectors, a focused electron-phonon coupling vertex selects a particular one ($\mathbf{q}_{\chi,1}$) of these (see Fig. 4 in Eiter et al.[27]). Here, we focus on the underlying susceptibility and its behavior as the $V_\pi$ ($\propto t_\perp$) is reduced. Extended Data Fig. 4a shows line cuts of the focused $\chi(\mathbf{q}, \omega = 0)$ along the (11) direction (where the electron-phonon coupling vertex is peaked) for several values of $V_\pi$. For both the local maxima, as $V_\pi$ is reduced, the peak in the susceptibility shifts towards lower $q$ values, which could explain the increase in CDW ordering wave-vector $q_c$ observed here – since $\mathbf{q}_c = \mathbf{c}^* - \mathbf{q}_\chi$. Extended Data Fig. 4b shows the obtained peak value as a function of $V_\pi$.

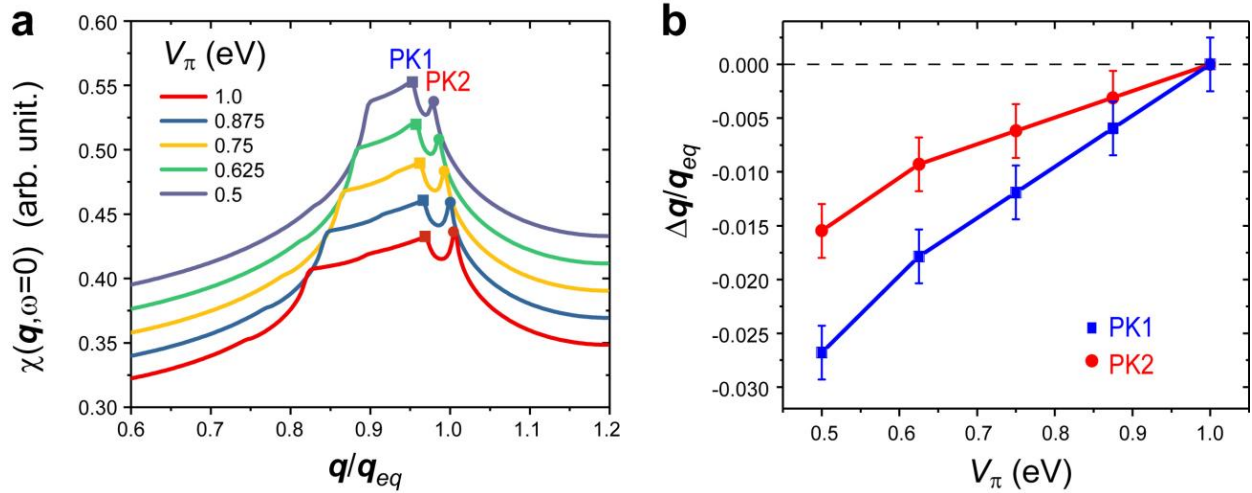

**Extended Data Fig. 4. Theoretical calculations of susceptibility ($\chi$) and nesting vector ($q_\chi$) | a,** $\chi(\mathbf{q}, \omega = 0)$ along $q_x = q_y$ (offset for clarity). Markers indicate the two local maxima in the susceptibility. The horizontal axis is scaled to the absolute maximum for $V_\pi = 1.0$ eV. **b,** Peak position of the charge susceptibility along the (11) direction. The error bars arise from the finite momentum resolution of the calculation.

To evaluate the theoretical ordering vector, we follow the methods outlined in Eiter et al.[27] The charge susceptibility is defined as

$$\chi(\boldsymbol{q},\omega) = 2\sum_{\alpha,\beta\in\pm}\sum_{\boldsymbol{k}}\frac{f\left(\epsilon_{\boldsymbol{k}+\boldsymbol{q}}^{\alpha}\right)-f\left(\epsilon_{\boldsymbol{k}}^{\beta}\right)}{\omega+\epsilon_{\boldsymbol{k}+\boldsymbol{q}}^{\alpha}-\epsilon_{\boldsymbol{k}}^{\beta}+i0^{+}}, \quad (M8)$$

where $f(x)$ is the Fermi function, $\epsilon_{\boldsymbol{k}}^{\alpha}$ is the bare dispersion for the $\alpha^{\text{th}}$ band at momentum $\boldsymbol{k}$. Both intra- and inter-band susceptibilities are included. The tight-binding model is based on the one used in Eiter et al. [27]. Specifically, at a 25% reduction in $V_\pi$ reported by ARPES, the CDW ordering vector $q_c$ should increase by ~1%. This is experimentally verified; at $F=1.85$ mJ/cm², a maximum shift of ≈2.5% is observed, whereas the lattice constant changes are negligibly small, within 0.03%; see Supplementary Data Fig. 2.

## 6. Microscale dynamics and thermalization: Three-temperature model

As was discussed by Tao, Han et al. [18,33], the energy exchanges between the microstates (electrons, phonons) can be described by a three-temperature model (3TM), which models after the photo-emission and ultrafast electron diffraction results in a self-consistent manner[18,32,39]. In this model, three effective thermal reservoirs are considered for energy exchanges: electrons, strongly coupled phonons, and weakly coupled phonons, with respective temperatures defined as $T_e$, $T_{\text{CDW, phonons}}$, and $T_{\text{lattice, phonons}}$. Using the coupling constants reported in Ref. [33], we extend the calculations to the specific laser fluences discussed here. The result for the nominal fluence of 1.86 mJ/cm² are presented in Extended Data Fig. 5, where the most relevant time scale is the relaxation time of hot carriers ($T_e$). We calculate it to be fluence-dependent, mainly due to an increase in the electronic heat capacity in the excited state, ranging from 2 to 5 ps over the applied $F$ range. This timescale serves to understand the ps decay of the emergent CDW state due to thermalization with the baths. This ps thermalization, which leads to lattice heating may be best seen in the phonon dynamics deduced from the Debye-Waller analyses presented in Fig. 4d. In this analysis, we show the CDW system reaches thermalization with the lattice modes after the metastable stage, where the system recovers to a thermal CDW state. The recovery time of the $c$-CDW order parameter $\tau_e$, as depicted in Fig. 3c, coincides with the thermalization timescale described in Fig. 4d, and here the timescale for the hot carrier relaxation.

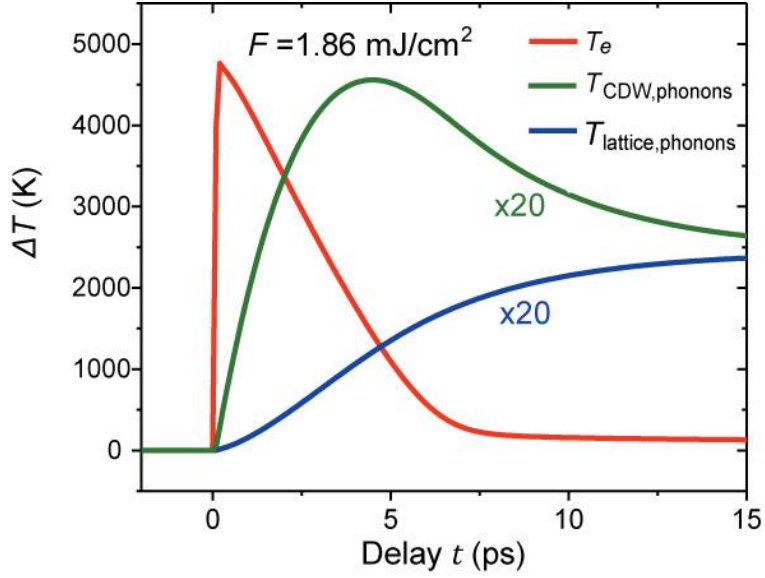

**Extended Data Fig. 5. Three-temperature model calculations of the temperature evolution | a,** The effective temperatures for the electronic bath ($T_e$) and the two lattice phonon baths (strongly coupled $T_{CDW, phonons}$ and weakly coupled $T_{lattice, phonons}$) at $F$=1.86 mJ/cm$^2$.

## 7. The Debye-Waller analysis

The relative changes introduced into the normalized structure factors of the lattice Bragg ($\hat{I}_{int,lattice}$) and CDW satellite ($\hat{I}_{int,CDW}$) peaks by a laser-induced suppression of CDW amplitude are considered here[19]. With a change $\Delta A$ in the lattice distortion, one obtains the complementary changes: $\hat{I}_{int,lattice} \approx 1 - 1/2(\boldsymbol{q} \cdot \Delta \boldsymbol{A})^2 sign(\Delta A) \approx e^{-1/2(\boldsymbol{q} \cdot \Delta \boldsymbol{A})^2}$ and $\hat{I}_{int,CDW} \approx 1/4(\boldsymbol{q} \cdot \Delta \boldsymbol{A})^2 sign(\Delta A) \approx 1 - e^{-1/4(\boldsymbol{q} \cdot \Delta \boldsymbol{A})^2}$. This latter intensity change represents a static contribution. In order to outline the (dynamic) contributions from the phonons ($\Delta u_{[hkl]}^2$), such a static factor must be deduced from the overall root-mean-squared displacement ($u_{hkl}$) deduced using the intensity evolution of the lattice Bragg peak (*hkl*): $\Delta u_{hkl}^2(t) = -1/q^2 \ln(\hat{I}_{int,lattice}) = \frac{1}{4}(\boldsymbol{q} \cdot \Delta \boldsymbol{A})^2 \times 2 + \Delta u_{[hkl]}^2(t)$ – a procedure referred to as Debye-Waller analysis. Here, an additional factor ½ in the static term is employed due to that CDW exists only in the Te nets, but not in the RTe layers. The static term is independently determined from $\hat{I}_{int,CDW}$.

# Supplementary Information

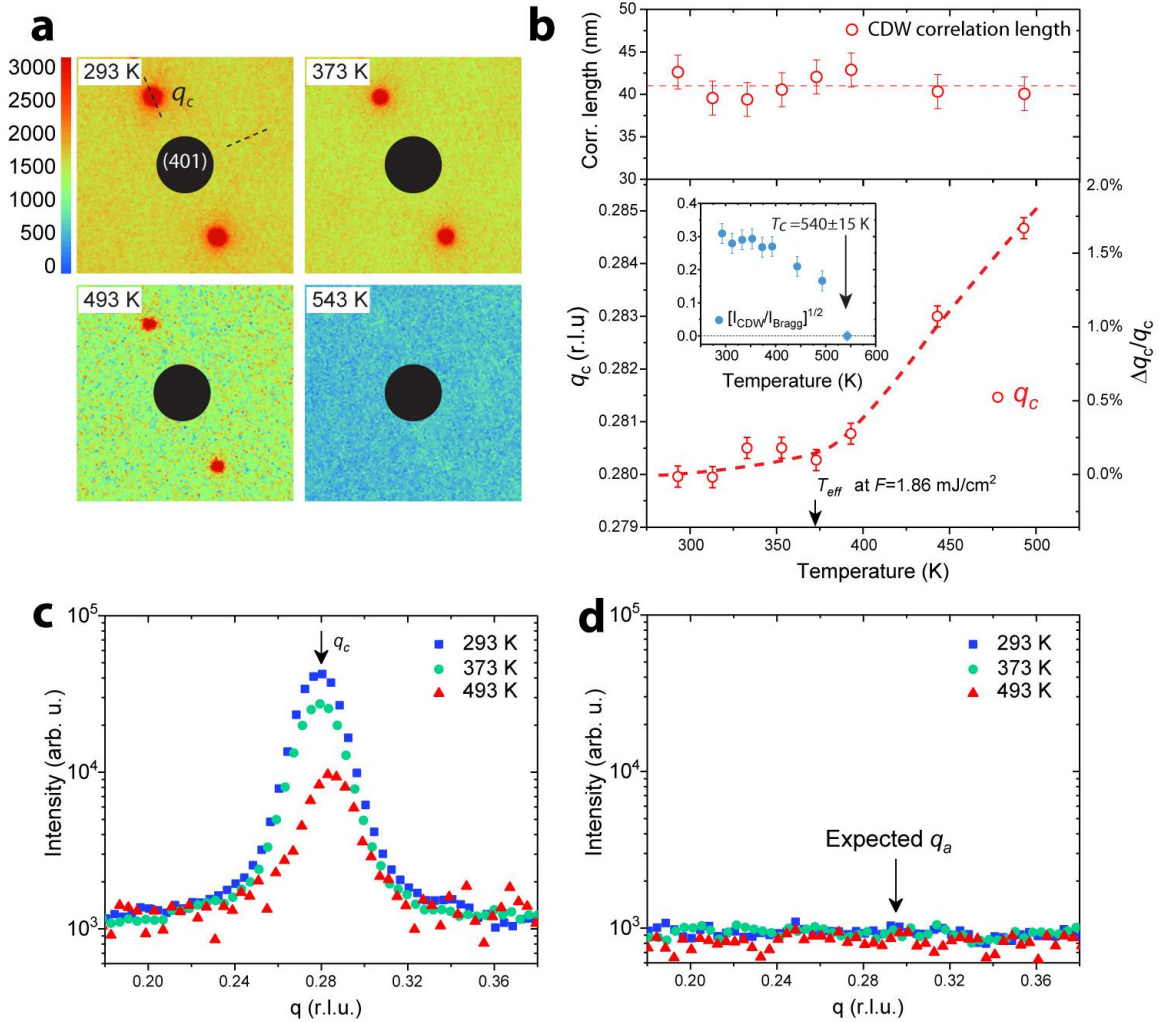

**Supplementary Fig. 1. Thermal phase transition of CDW state in CeTe$_3$ | a.** The diffraction images showing the CDW satellite peaks evolution at selected temperatures near phase transition obtained from a transmission electron microscope (JEOL 2010F) using the Gatan OneView 4k×4k camera. **b.** Top panel shows the CDW correlation length evolution as the temperature is increased from room temperature. The bottom panel shows the corresponding shift of the CDW wav-vector ($q_c$). The inset shows the corresponding relative changes in the CDW satellite intensity, approaching the melting temperature $T_c$=540±15 K. **c &d.** The line profiles along the $q_c$ and $q_a$ (expected) directions at three selected temperatures.

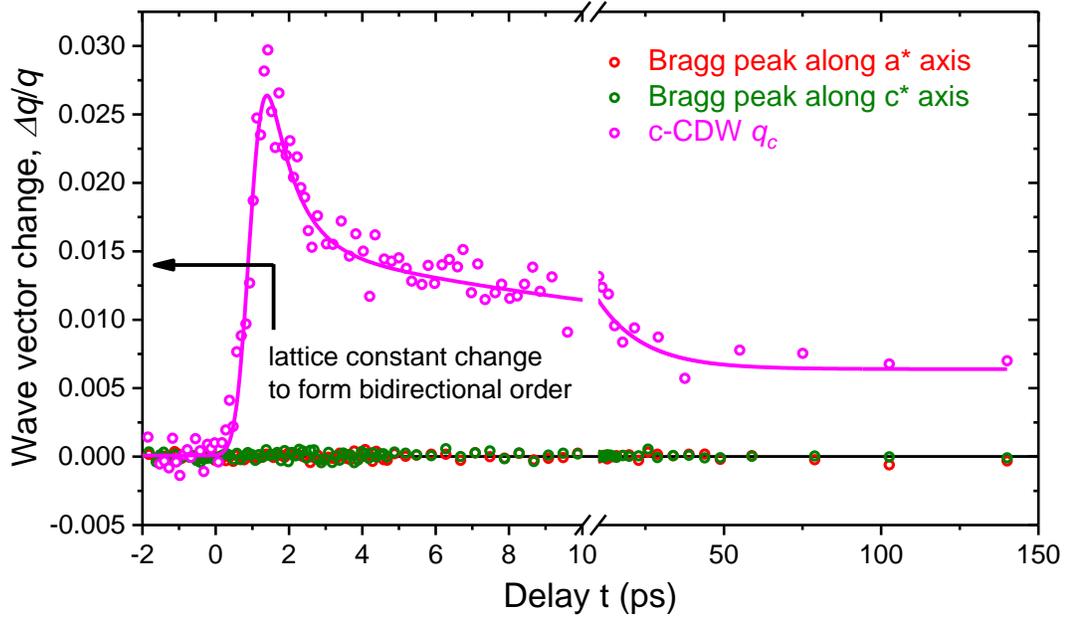

**Supplementary Fig. 2. The relative wave-vector shifts of the CDW and lattice peaks in photo-induced phase transition.** The Bragg peak/lattice constant changes are based on the (600) and (006) reflections. The arrow marks the change needed for forming the bi-directional order based on the lattice constant changes from $CeTe_3$ to the neighboring $TbTe_3$ where such order starts to emerge thermodynamically at the low temperature (see Fig. 1c).

23